\documentclass[12pt,preprint]{aastex}

\usepackage{natbib}
\usepackage{latexsym,amssymb,verbatim,amsmath}

\slugcomment{Accepted by ApJ, May 2016}
\shorttitle{Hydrides in Diffuse Clouds}
\shortauthors{Awad et al.}

\begin{document}


\title{On the chemistry of hydrides of N atoms and O$^+$ ions}


\author{Zainab Awad$^{1,2}$}
\affil{$^1$Astronomy, Space Science, \& Meteorology Department, Faculty of Science, Cairo University, Giza , Egypt \\
$^2$Physics and Astronomy Department, University College London, Gower Street, London WC1E 6BT, UK}
\email{zma@sci.cu.edu.eg}
\and
\author{Serena Viti$^2$ and David A. Williams$^2$} 
\affil{$^2$Physics and Astronomy Department, University College London, Gower Street, London WC1E 6BT, UK}




\begin{abstract}
Previous work by various authors has suggested that the detection by Herschel/HIFI of nitrogen 
hydrides along the low density lines of sight towards G10.6-0.4 (W31C) cannot be accounted for by 
gas-phase chemical models. In this paper we investigate the role of surface reactions on dust grains 
in diffuse regions, and we find that formation of the hydrides by surface reactions on dust grains with efficiency 
comparable to that for H$_2$ formation reconciles models with observations of nitrogen hydrides. 
However, similar surface reactions do not contribute significantly to the hydrides of O$^+$ ions detected by Herschel/HIFI 
present along many sight lines in the Galaxy. The O$^+$ hydrides can be accounted for by conventional 
gas-phase chemistry either in diffuse clouds of very low density with normal cosmic ray fluxes or in 
somewhat denser diffuse clouds with high cosmic ray fluxes. Hydride chemistry in dense dark clouds 
appears to be dominated by gas-phase ion-molecule reactions.

\end{abstract}


\keywords{Astrochemistry - ISM: abundances, clouds, molecules - Dust}

\section{Introduction}
\label{intro}
Observations using Herschel/HIFI have been made of rotational lines of 
hydrides NH, NH$_2$, and NH$_3$ of nitrogen atoms towards G10.6-0.4 \citep{per10} 
and of the hydrides OH$^+$, H$_2$O$^+$, and H$_3$O$^+$ of oxygen ions towards the same object \citep{ger10} 
and towards Orion KL \citep{gup10}. 
The nitrogen hydrides have also been observed towards W49 \citep{per12} 
and towards the Class 0 protostar IRAS 16293-2422 \citep{hily10}. 
The oxygen ion, O$^+$, hydrides have been studied along 20 lines of 
sight and in many velocity components towards bright submillimetre 
continuum sources in the Galaxy \citep{indr15}. 
Hydrides not directly associated with stellar sources are generally believed to be 
present in relatively low density gas along the line of sight.

These detections have stimulated a re-examination of chemistry of hydrides 
in both diffuse (e.g. \citealt{holn12, indr15}) and 
dark (\citealt{disl12, legal14, legal14sf}) interstellar clouds. 
Both gas-phase and grain surface reactions have been invoked in these 
studies.  

In this paper, we focus mainly on diffuse gas because of its relative simplicity. We 
re-examine the predictions of hydride abundances from 
diffuse cloud models based on gas-phase chemistries to which surface 
reactions may also contribute. In particular, we study the sensitivity of 
the chemistry to variations in the physical parameters that control the 
chemistry. We shall examine first the chemistry of the nitrogen hydrides, 
and shall conclude, along with other authors, that gas-phase schemes are 
incapable of supplying the abundances implied by the observations of 
nitrogen hydrides in diffuse gas. However, grain surface reactions with plausible 
efficiencies are capable of supplying nitrogen hydrides in the observed 
abundances and with the observed ratios of NH$_3$/NH$_2$ and NH$_3$/NH. 
We shall find that the reverse is true for the hydrides of O$^+$, 
i.e., that reactions of O atoms with grain surfaces contributing H$_2$O molecules to the 
gas-phase do not significantly affect the O$^+$ hydrides; 
they do, however, slightly affect the neutral hydrides OH and H$_2$O.

\section{Surface chemistry}
\label{schem}
Hydride formation in surface reactions was first discussed over half a 
century ago \citep{mccr60}. 
\citet{pic77} explored an 
extensive nitrogen chemistry in a ``gas-phase + grain surface" reaction 
model. Very detailed studies of gas + grain chemistry in diffuse clouds on sight lines 
towards $\zeta$ Per and $\zeta$ Oph were made by \citet{wag93MN}. 
In particular, a good match for more than 30 atomic and 
molecular abundances towards $\zeta$ Per was obtained, including NH, for 
which the column density had been measured by \citet{mey91}. 
The NH column density predicted by \citet{wag93MN} 
for the $\zeta$ Oph model was later confirmed by ultra-high resolution observations \citep{craw97}. 
Gas-phase models fail to supply enough NH for these 
lines of sight, so these computations may be regarded as strong support 
for a contribution to interstellar chemistry from surface reactions in diffuse clouds.

Several experimental studies have been made of sequential hydrogenation of nitrogen atoms in various 
matrices and ice analogues (\citealt{hir95,hid11,fed15}; see also the review by \citealt{lin15}). 
These studies confirm that this hydrogenation is efficient and probably 
barrier-free even at the low temperatures of interstellar dust grains. A detailed description of the process 
is not possible because of the very limited data on binding energies, mobilities, and desorption processes 
of the various species involved. In this work, we adopt the version of the \citet*{wag96} surface chemistry in 
which a neutral nitrogen or oxygen atom arriving at a grain surface is 
converted directly to NH$_3$ or H$_2$O and ejected promptly into the gas; the 
desorption is assumed to be driven by energies released in the reaction. 
The reaction is assumed to occur with the same efficiency as the H$_2$ 
formation reaction, but the arrival rate of N or O atoms at grain surfaces 
is reduced by the square root of the atomic mass of the atom, and, of 
course, by the abundance of atoms relative to hydrogen (all values of 
which are taken from \citet{asp09} who give the fractional abundances of 
N and O relative to hydrogen to be 6.761 $\times$ 10$^{-5}$ and 4.898 $\times$ 10$^{-4}$; respectively). 

The H$_2$ formation rate adopted in our computations is 
1.3 $\times$ 10$^{-17}~ T^{1/2}~ n$(H)$~ n_{\text{H}}$ in cm$^{-3}$~s$^{-1}$ (e.g. \citealt{jura74,tiel05}). 
Similarly, the rate of formation of N- and O-hydrides can be expressed, in cm$^{-3}$~s$^{-1}$, as
3.5 $\times$ 10$^{-18}~T^{1/2}~ n$(N)$~ n_{\text{H}}$ and 3.3 $\times$ 10$^{-18}~T^{1/2}~ n$(O)$~ n_{\text{H}}$; 
where $T$ is the gas kinetic temperature, $n$(X) is the number density of free X atoms, and $n_{\text{H}}$ is 
that of hydrogenin any form.

\section{Cloud model and gas-phase chemistry}
\label{mod}

The cloud geometry is a one-dimensional slab of uniform density, 
illuminated by the interstellar radiation field from one side. The 
reference model ({\it hereafter}, RM) has $n_{\text{H}}$ = 100 cm$^{-3}$, is illuminated by the Draine 
interstellar radiation field (RF), here denoted by G$_0$, and irradiated by 
cosmic rays that cause hydrogen ionization at a rate $\zeta_{\text{ISM}}$ = 1.3 $\times$ 
10$^{-17}$ s$^{-1}$ \citep{wak05}. 
The gas:dust ratio is assumed uniform and such that one visual magnitude of 
extinction is equivalent to 1.6 $\times$ 10$^{21}$ total H cm$^{-2}$ \citep{dra11book}. The model results 
described below are normally for a cloud of one visual magnitude.

The gas-phase chemistry that leads to the formation of nitrogen hydrides 
in dense dark clouds has been reviewed by \citet{legal14sf}. 
In these clouds, nitrogen is mainly in N atoms and N$_2$ molecules. Exchange reactions 
of N and N$_2$ with H$_2$ molecules are strongly suppressed at the low 
temperatures in diffuse and dark clouds. Hence, these neutral routes with 
hydrogen molecules do not proceed. The reaction of N atoms with H$_3^+$ ions 
(created by cosmic ray ionization of H$_2$) has a high activation energy; in 
fact, the reverse reaction H$_2$ + NH$^+ \longrightarrow$ H$_3^+$ + N is a fast proton transfer. 
The product N$_2$H$^+$ formed in the proton transfer reaction N$_2$ + H$_3^+ \longrightarrow$ N$_2$H$^+$ + 
H$_2$ has a minor channel of dissociative recombination leading to NH that 
is important in dark dense clouds, but is unimportant in diffuse clouds 
because in them N$_2$ is a very minor component. For example, \citet{kna04}, 
using far UV absorption spectroscopy of matter in diffuse clouds 
towards HD 124314, was able to show that only about 10$^{-4}$ of nitrogen was 
in the form of N$_2$.

In diffuse clouds the main route to form NH$^+$, and by subsequent hydrogen 
abstraction reactions and dissociative recombinations, to neutral nitrogen 
hydrides is N$^+$ + H$_{2} \longrightarrow$ NH$^+$ + H, where the N$^+$ ion 
is formed by cosmic ray ionization of N atoms. This 
rate coefficient, including its dependence on the H$_2$ ortho/para ratio, has 
been the subject of several theoretical and experimental investigations (e.g. 
\citealt{lebo91,marq88,gerl93,disl12,zym13}).
The experimental results of Zymak et al. confirm 
the rate coefficient of Dislaire et al. However, the role of fine 
structure excitation in N$^+$ ions in the reaction may require further 
investigation. We adopt the value of rate coefficient for the N$^+$ + H$_2$ 
reaction given in the UDfA\footnote{Also known as UMIST; http://udfa.ajmarkwick.net/} database \citep{mce013}. At 100 K, this 
differs from that of Zymak et al., by a factor of three or less, and from 
the value in the KIDA database by a factor of less than 2. The rate 
coefficients adopted for the main reactions determining nitrogen hydrides 
are shown in Table \ref{tab:1}. 

The gas-phase chemistry that leads to the formation of hydrides of O$^+$ ions 
has been reviewed recently by \citet{holn12} 
and by \citet{indr15}. 
In their Figure 1, Hollenbach et al. summarize the two main 
channels to the ions OH$^+$, H$_2$O$^+$, and H$_3$O$^+$. The first of these is 
initiated by the cosmic ray ionization of atomic hydrogen, followed by the 
accidental resonance charge exchange of the hydrogen ion with an oxygen 
atom, forming O$^+$ which then successively abstracts H from H$_2$ to form OH$^+$, 
H$_2$O$^+$, and H$_3$O$^+$. Dissociative recombination of these ions leads to neutral 
products O, OH, and H$_2$O. All these reactions must compete with 
photoprocesses and other reactions. This first channel is significant in 
diffuse clouds, where both atomic and molecular hydrogen are abundant. 

\begin{figure*}
\begin{center} 
\includegraphics[width=12cm]{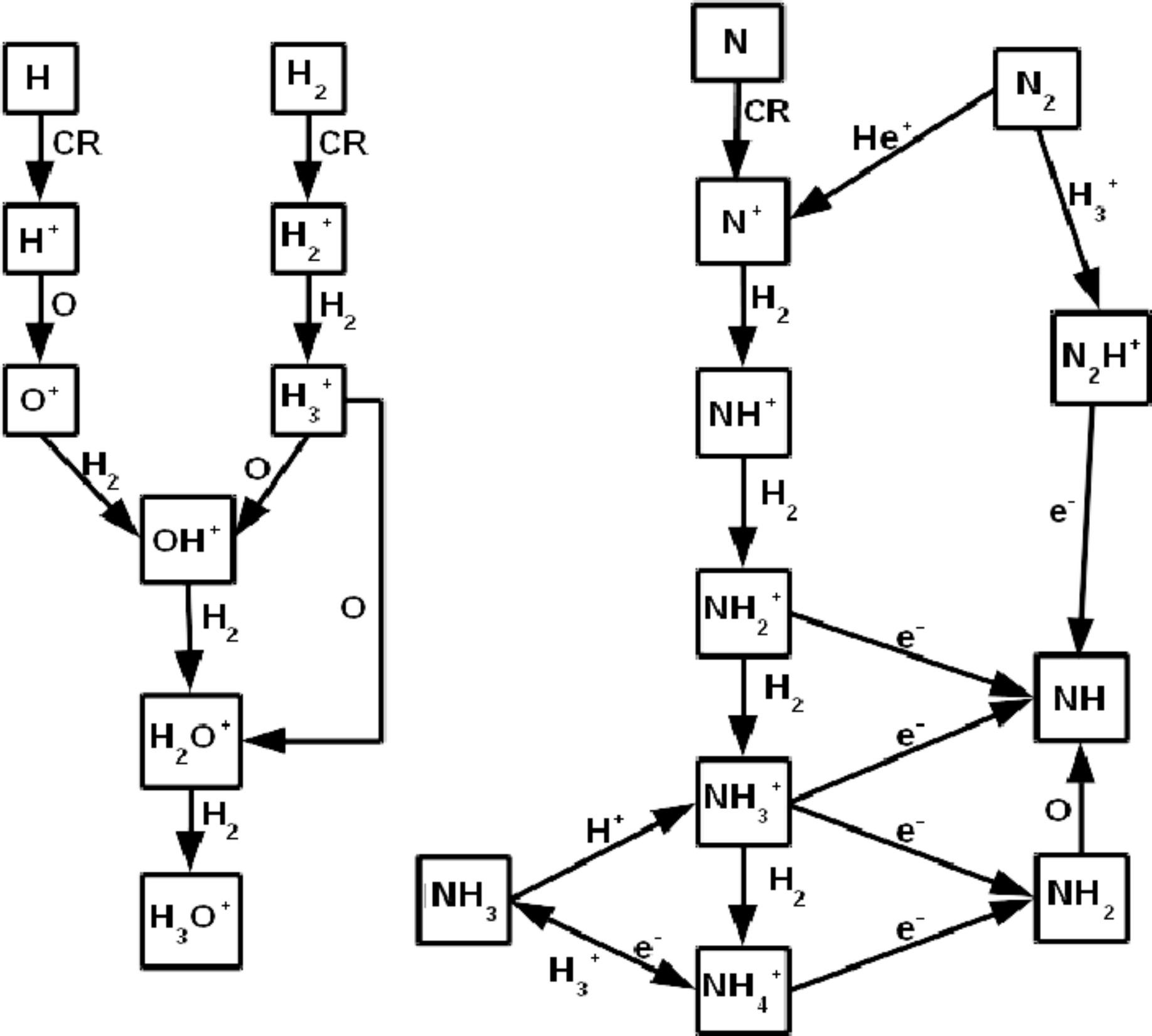} 
\caption{The main routes in the gas-phase chemistry forming hydrides of N and O$^+$. Diagrams modified from \citet{holn12} and \citet{legal14sf}.}
\label{fig:0}
\end{center}
\end{figure*}

The second channel involves the reaction of O atoms with H$_3^+$ ions arising 
from the cosmic ray ionization of H$_2$. The reaction generates OH$^+$ and H$_2$O$^+$ 
ions which, as in the first channel, abstract hydrogen from H$_2$ molecules 
to form the full set of O$^+$ hydrides. This channel is likely to be dominant 
in denser regions of the interstellar medium. A sketch of the main gas-phase entry 
chemical reactions in our model to form N-, O-, and O$^+$-hydrides is illustrated in 
Figure \ref{fig:0}, and the main reactions are listed in Table \ref{tab:1}.

Rate constants are calculated using the mathematical expression described 
in UMIST database releases (e.g. \citealt{milr97,lete2000,woo07,mce013}) as follows:
$$k = 
\begin{cases} 
\mbox{a) if two-body reaction}\\
\alpha ~ (T/300)^{\beta} ~ exp(-\gamma/T)~~ \mbox{cm$^3$ s$^{-1}$} \\ \\
\mbox{b) if CR ionization reaction, for $\zeta = \zeta_{\text{ISM}}$}\\
\alpha \quad\quad \mbox{s$^{-1}$}
\end{cases}
$$ %
where $\zeta $ is the CR ionization rate, and the constants $\alpha$, $\beta$, and $\gamma$ are 
listed in Table \ref{tab:1}. 

The elements included in the chemistry are H, He, O, C, N, S, Mg, Si, and 
Cl, and the chemical network includes 158 species interacting in 
2132 chemical reactions whose rate coefficients are taken from the 
UDfA 2012 database \citep{mce013}. 
The rate equations for all species are integrated and molecular abundances provided as a function of time at a 
particular visual extinction, $A_V$. A full description of the chemical model, UCL$\_$CHEM, used in the 
current work is given in \citet{viti99}, with some updates described in \citet{viti04} and \citet{rob07}.

\section{Nitrogen Hydrides}
\label{nh}
\begin{figure*}
\begin{center} 
\includegraphics[width=16cm]{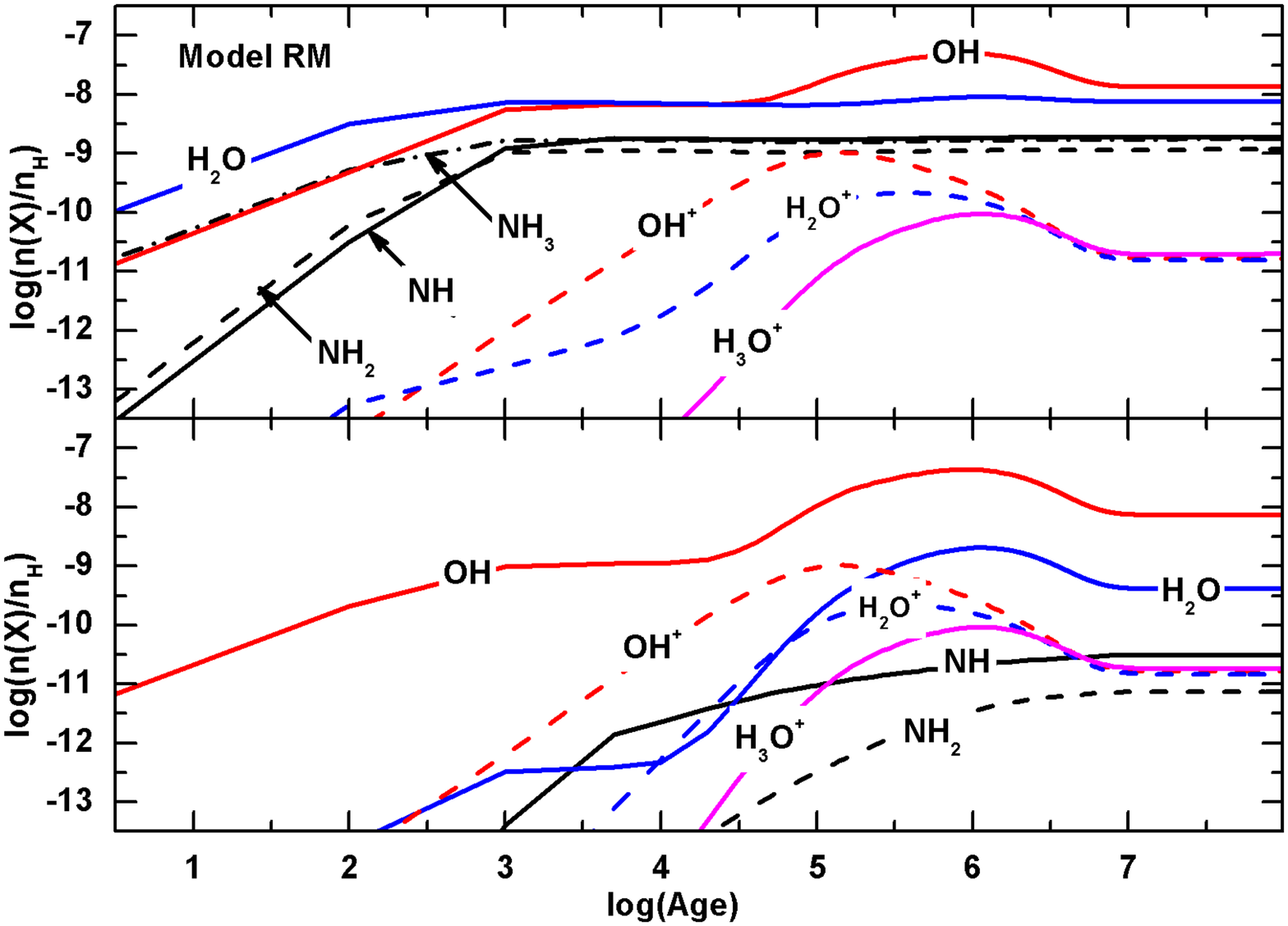} 
\caption{Fractional abundances in the Reference Model ($n_{\text{H}}$ = 100 cm$^{-3}$, 
$\zeta$ = 1 $\times~ \zeta_{\text{ISM}}$, RF = 1 $\times$ G$_0$, but with H$_2$(t=0)= 0) showing the time 
dependence of all the hydrides, with surface reactions included (top) and 
excluded (bottom). Note that when surface reactions are excluded, the fractional abundance of NH$_3$ 
is less than 10$^{-13}$ (at steady-state, its value is $\sim$ 5$\times$10$^{-15}$).}
\label{fig:1}
\end{center}
\end{figure*}

Figure \ref{fig:1} shows that the fractional abundances at steady-state of the three 
nitrogen hydrides for the RM are all larger than 10$^{-9}$ when grain surface reactions 
contribute, and at least an order of magnitude lower when they are excluded. 
Figure \ref{fig:1} shows that, in the RM, the NH$_3$ fractional abundance is less than 10$^{-13}$ and 
probably undetectable, when excluding grain surface contribution. These steady-state fractional abundances are 
- as expected - found to be independent of the cosmic ray ionization rate (for a range up 
to 50 times the canonical interstellar value) and also independent of the initial H$_2$ fraction. 

The fractional abundances depend on the cloud density; see Figure \ref{fig:2}a. The nitrogen hydride 
fractional abundances are also sensitive to the radiation field; see Figure \ref{fig:2}b. Note that 
if surface reactions are excluded, then the pure gas-phase chemistry generates fractional abundances of 
nitrogen hydrides that are approximately proportional to the cosmic ray ionization rate.

\section{Oxygen atom and ion hydrides}
\label{oh}

Figure \ref{fig:1} shows results in the RM for all hydrides 
of O atoms and O$^+$ ions, both when surface reactions contribute and when 
they do not. There is a significant time-dependence in the chemistry. This 
arises because the chemistry is driven by slow cosmic ray ionization, and 
also because the chemistry is affected by the H/H$_2$ balance which takes a 
long time ($\sim$ several $\times$ 10$^9$ yr /$n_{\text{H}}$ cm$^{-3}$) to come to steady-state in 
diffuse clouds. In fact, the time to attain steady-state may exceed 
the lifetime of a diffuse cloud (often taken to be a few million years). 
In that case, chemistry of O$^+$ ions may never reach steady-state. 

Figure \ref{fig:1} indicates a very slight enhancement in OH and H$_2$O RM fractional abundances 
when surface reactions are operating. However, surface reactions do not 
affect the RM fractional abundances of the O$^+$ hydrides either at the peak fractional abundances 
(at $\sim$ 10$^5$ yr) or at steady-state. The peak fractional abundances of the 
ions are on the order of 10$^{-10}$ in the RM, while their steady-state values 
are an order of magnitude lower.

The O$^+$ hydrides are found to be fairly insensitive to the radiation field; 
this is because their chemistry is dominated by fast ion-molecule 
reactions. However, the fractional abundances of these hydrides are sensitive to both gas density and 
to the cosmic ray ionization rate, see Figure \ref{fig:3}a and \ref{fig:3}b.

Figure \ref{fig:3}a shows that the O$^+$ hydride fractional abundances depend approximately 
inversely on the density. For example, with $n_{\text{H}}$ = 10 cm$^{-3}$, peak fractional abundances 
(i.e. all ion abundances measured at the OH$^+$ peak) are larger by an order 
of magnitude (and arise earlier, at an evolutionary time $\sim$ 10$^5$ yr) than 
for $n_{\text{H}}$ = 100 cm$^{-3}$, (where peak fractional abundances arise at $\sim$ 10$^6$ yr). Similar 
statements can be made for the steady-state fractional abundances. The effects of 
high cosmic ray ionization rates on peak abundances are shown in Figure \ref{fig:3}b. 
Abundances are roughly proportional to the ionization rate. Such behavior has been described 
by \citet{holn12} and was obtained in earlier work by \citet{bay11}.

\section{Discussion and Conclusions}
\subsection{Nitrogen hydrides in diffuse clouds} 

It is clear from results from the RM (see Figure \ref{fig:1}) that gas-phase models 
fail to provide nitrogen hydrides in the required abundances as measured 
towards G10.6-0.4 by \citet{per10}, 
who came to the same conclusion. The data of \citet{indr15} show that this gas is mainly 
atomic and is therefore quite diffuse. In particular, pure gas-phase models such as that of 
\citet{lepet04}, cannot provide enough NH$_3$ in diffuse clouds to satisfy the NH$_3$/NH and 
NH$_3$/NH$_2$ ratios obtained from observations (see below). In gas-phase models of low density 
clouds, the predicted ratios fail by several orders of magnitude. However, the contribution from 
surface reactions can be significant and comparable to the observed fractional  abundances. In 
the (arbitrarily chosen) RM almost the same amount of nitrogen is in the three nitrogen 
hydrides as is observed to be the case towards G10.6-0.4.

\begin{figure*}
\begin{center} 
\includegraphics[trim=1cm 1cm 1cm 1.2cm, clip=true,width=15cm]{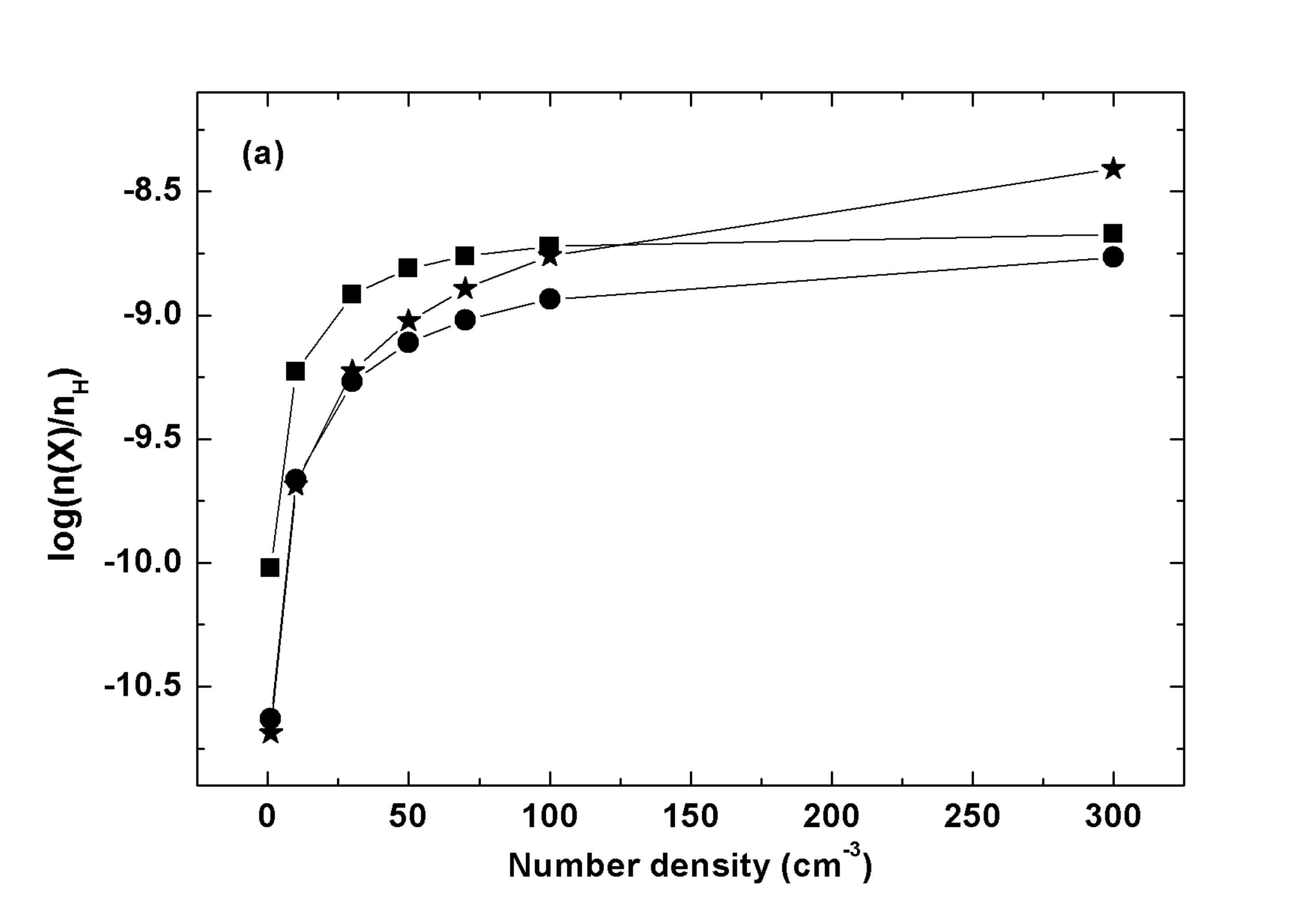} 
\includegraphics[trim=1cm 0.5cm 1cm 1cm, clip=true,width=15cm]{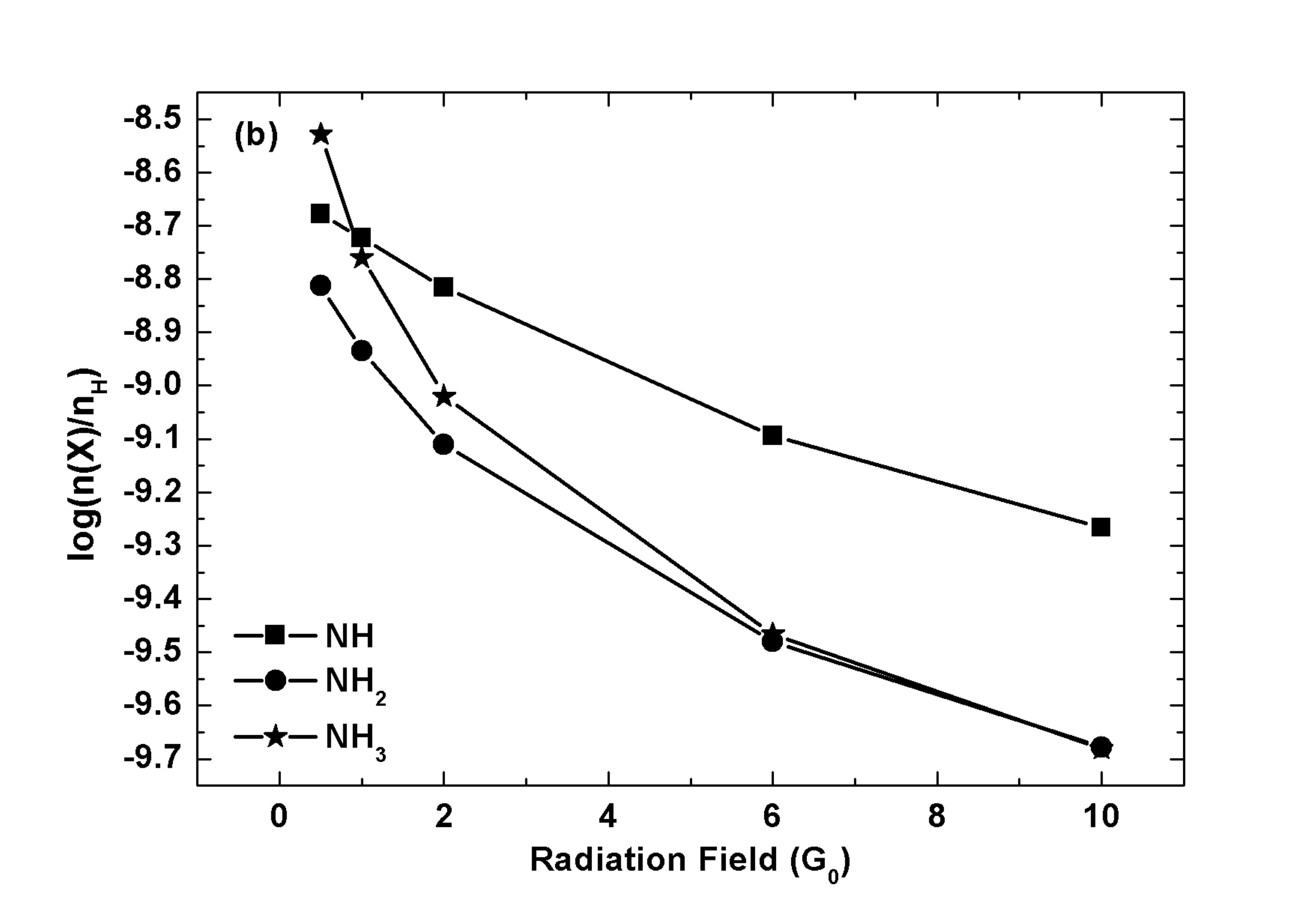} 
\caption{(a) Steady-state fractional abundances of NH, NH$_2$, and NH$_3$ for 
number densities between $n_{\text{H}}$ = 1 and 300 cm$^{-3}$. (b) As for (a), but for 
radiation fields RF = 0.1 - 10 $\times$ G$_0$. }
\label{fig:2}
\end{center}
\end{figure*}

\begin{figure*}
\begin{center} 
\includegraphics[trim=1cm 1cm 1cm 1.2cm, clip=true,width=15cm]{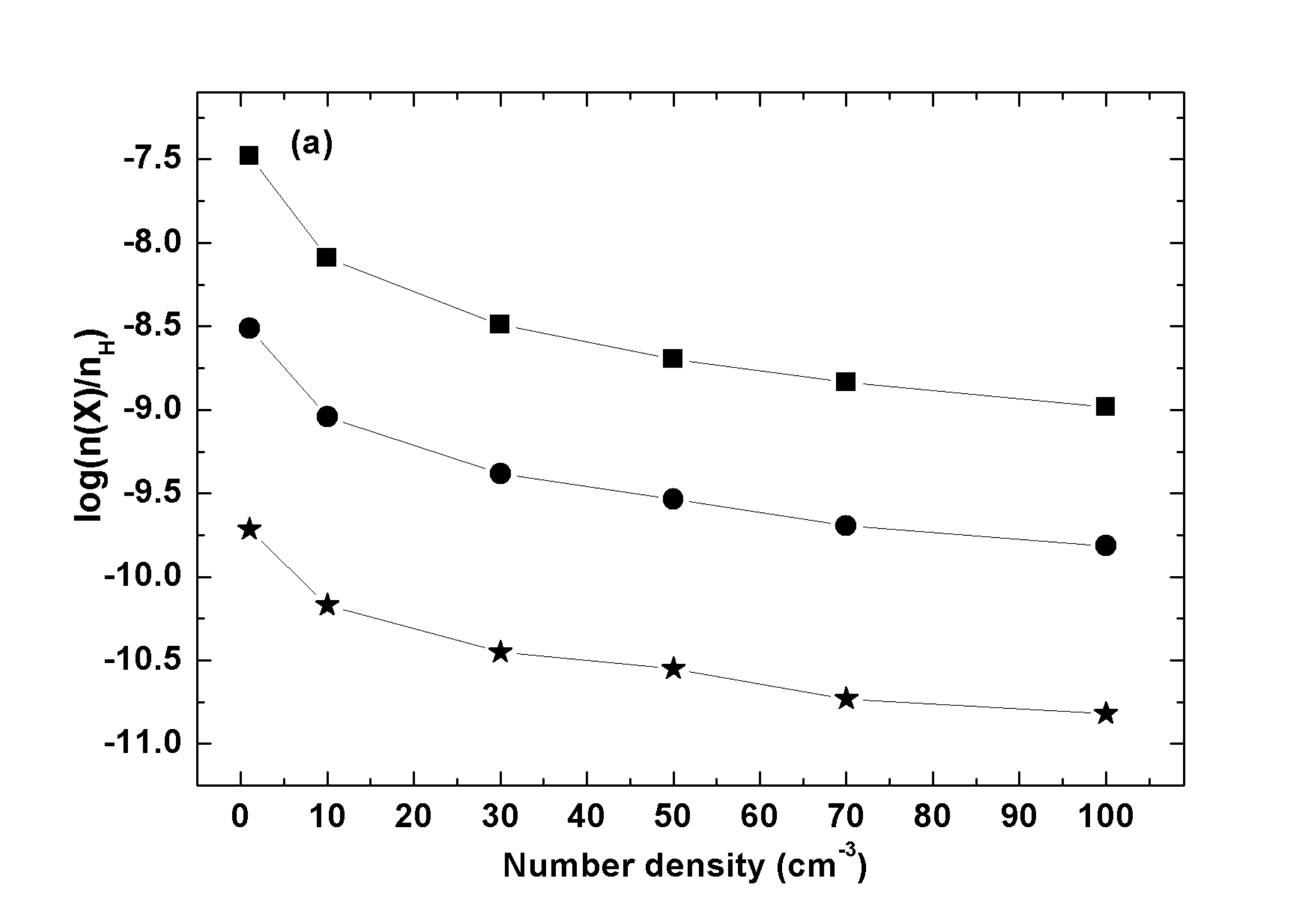} 
\includegraphics[trim=1cm 0.5cm 1cm 1cm, clip=true,width=15cm]{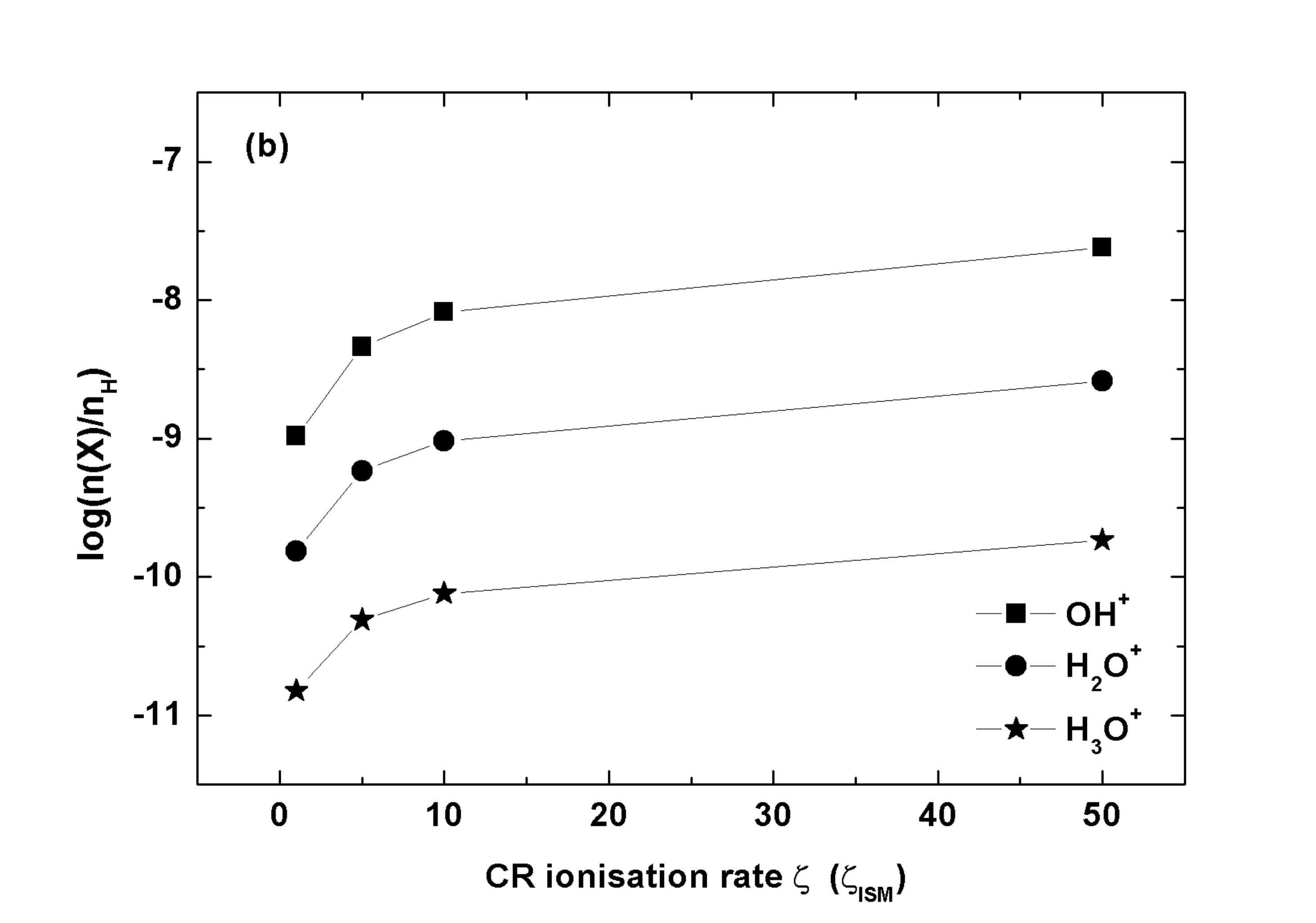} 
\caption{As for Figure \ref{fig:2}, but for OH$^+$, H$_2$O$^+$, H$_3$O$^+$, with fractional abundances 
at peak OH$^+$. (a) is for cloud number densities $n_{\text{H}}$ = 1 - 100 
cm$^{-3}$. (b) is for cosmic ray fluxes 1 - 50 $\times ~ \zeta_{\text{ISM}}$.}
\label{fig:3}
\end{center}
\end{figure*}

However, the distribution of nitrogen among the three hydrides does not 
match the observed results towards G10.6-0.4. \citet{per10} found the fractional abundances for 
NH, NH$_2$, and NH$_3$ to be approximately (6, 3, 3)$\times$ 10$^{-9}$, respectively. 
Figures \ref{fig:2}a and \ref{fig:2}b show that the three species respond differently to changes in 
cloud density and radiation field. The computed ratios NH$_3$/NH and NH$_3$/NH$_2$ may be compared 
with the observed values of 0.5 and 1.0, respectively, see Table \ref{tab:2}. 
Models with low number densities ($n_{\text{H}} \sim$ 30 cm$^{-3}$) and radiation fields (0.5 - 1 G$_0$) 
can match the observed ratios, but the total amount of nitrogen in all 
nitrogen hydrides is less than a fifth of that observed. Evidently, we 
need somewhat higher number densities (say, $n_{\text{H}} \sim$ 100 cm$^{-3}$) with somewhat 
reduced radiation field ($\sim$ 0.25 G$_0$) to have the observed total nitrogen 
hydride fractional abundance ($\sim$ 1 $\times$ 10$^{-8}$). For the RM, the ratios do not match 
those observed. Figures \ref{fig:2}a and \ref{fig:2}b show that it is not possible 
simultaneously to match both high fractional abundances and appropriate ratios of the 
nitrogen hydrides detected along the line of sight towards G10.6-0.4.

The simplest resolution of this mismatch to which we are forced is, 
therefore, to reconsider the initial assumption which was that nitrogen 
atoms arriving at grain surfaces are all converted to NH$_3$ molecules and 
injected into the gas-phase. We show in Table \ref{tab:4} our results 
for the RM in which the nitrogen hydrides ejected into the gas-phase are 
distributed equally between NH, NH$_2$, and NH$_3$, rather than all in NH$_3$. With 
this modification, our calculations (see Table \ref{tab:4}, Model A) for the RM but with a 
reduced radiation field (0.25 G$_0$) give ratios NH$_3$:NH and NH$_3$:NH$_2$ of 0.5 
and 0.9, close to observed, and with fractional abundances that are only slightly 
lower than are observed, a deficiency that can be resolved at a slightly higher density 
(Table \ref{tab:4}, Model B). 

\citet{per12} have re-analyzed the G10.6-0.4 nitrogen hydride data, and also present data for foreground diffuse 
material towards G49N. The column density ratios $N$(NH)/$N$(o-NH$_3$) and $N$(o-NH$_2$)/$N$(o-NH$_3$) over different 
velocity components are fairly constant at 3.2 and 1.9, respectively, for W49N and 5.4 and 2.2 for G10.6-0.4. The 
fractional abundance of o-NH$_3$ is $\sim$ 2 $\times$ 10$^{-9}$ in both sources. These results are better fitted by models 
such as those shown in Figure \ref{fig:2}b. The material in nitrogen hydrides is $\sim$ 50\% larger in 
G10.6-0.4 than in W49N. This variation could be accounted for either by a slightly lower gas density 
or a larger radiation field in W49N (see Figures \ref{fig:2}a, and \ref{fig:2}b).

\subsection{Nitrogen hydrides in dense gas} 

The main purpose of this paper has been to discuss the formation of 
hydrides of N atoms and O$^+$ ions in diffuse gas, such as that towards 
G10.6-0.4. However, it is useful to make a comparison with models of 
chemistry that have been developed to account for the abundances of 
nitrogen hydrides and their ratios in much denser gas. An exceptionally 
comprehensive study of the gas-phase chemistry of nitrogen hydrides in 
dark clouds has been carried out by \citet{legal14}. 
The chemical network used in the present study is similar but significantly larger than 
that of Le Gal et al., but the Le Gal et al. study is very much more 
detailed in several important respects. In our study, the effects of the 
spin symmetries of the nitrogen hydrides and the ortho-para forms of 
molecular hydrogen have been ignored, whereas Le Gal et al. have 
determined self-consistently their role in the chemistry. Le Gal et al. 
have also demonstrated that the amounts of carbon, oxygen and sulfur 
available in the gas have significant effects on the abundances of 
nitrogen hydrides. These occur through the enhancement of N$_2$ through 
neutral chemistries such as N + CH $\longrightarrow$ CN + H followed by 
CN + N $\longrightarrow$ C + N$_2$; and OH + N $\longrightarrow$ NO + 
H followed by NO + N $\longrightarrow$ N$_2$ + N. The role of sulfur, as 
S$^+$, is to destroy CH and (marginally) OH, and therefore inhibit the growth 
of N$_2$. All these reactions are included in the network of the present 
paper, and so the general behaviour of chemistry is similar in both 
models, even though the effects of spin symmetries and H$_2$ ortho/para 
reactions have been ignored in our work. Given sufficient time (typically, 
several million years) gas-phase reactions convert most of the available 
nitrogen to N$_2$ molecules, and reactions N$_2$ + H$_3^+ \longrightarrow$ 
N$_2$H$^+$ + H$_2$ followed by the dissociative recombination of N$_2$H$^+$ 
(discussed in Section 3, above) become the major source of NH and hence 
of other nitrogen hydrides.

\citet{legal14sf} have compared the predictions of their models with 
observational results obtained by Herschel/HIFI of NH, NH$_2$, and NH$_3$ 
detections in the envelope of the protostar IRAS 16293-2422. They also 
reassessed earlier determinations of the hydride abundances by \citep{hily10} 
and obtained the ratios NH:NH$_2$:NH$_3$ = 3:1:19, in a cloud in 
which $N_{\text{H}}$ = (3.0 $\pm$ 1.5) $\times$ 10$^{22}$ cm$^{-2}$. Le Gal et 
al. find that their model gives good fits to both abundances and ratios of 
nitrogen hydrides with specific choices for C/O and S abundances.

We come to a similar conclusion for our model with more extensive but less 
detailed gas-phase chemistry. Without detailed fitting, our purely gas-phase 
chemistry for a cloud with density $n_{\text{H}}$ = 10$^{4}$ cm$^{-3}$ 
and $A_V$ = 3 mag leads to the nitrogen hydride ratio 2:1:8, reasonably similar to the 
revised observational values. Our model for gas-phase + grain surface 
chemistries gives a much worse fit to the ratio.

Thus, it appears that the role for surface reactions in determining 
nitrogen hydride abundances is clear in diffuse clouds, but not in 
dense dark clouds. In dark clouds, the high abundance of N$_2$ enables the 
opening up of an additional and efficient gas-phase channel leading to nitrogen hydrides. 

The situation in star-forming regions is, of course, more complicated 
than in quiescent dark clouds. The timescale for freeze-out in clouds of 
number density $n_{\text{H}}$ = 10$^{4}$ cm$^{-3}$ is much less than the 
likely cloud age. Molecules are very likely to visit a grain surface, and depending on 
the timescale for desorption may make such visits many times. Thus, the simple picture 
presented here may, in fact, be more complicated. For example, models of complex chemistry, 
such as that of \citet{garod08} assume that N, NH, and NH$_2$ are hydrogenated in grain 
surface chemistry in regions of star formation. Such processes would 
affect the hydride abundances and ratios.

\subsection{Hydrides of oxygen ions} 

There is a wealth of observational data on the hydrides of O$^+$ observed on 
20 galactic lines of sight towards bright submillimetre continuum sources \citep{indr15}. 
On each line of sight, a number (from 2 to 13) of 
distinct velocity components are found, so that the total number of 
sources is over 100. Positive detections of the ion H$_3$O$^+$ were made on 7 of 
the 20 lines of sight and 16 of the velocity components (most of these 
towards Sgr B2 (M and N) where the ions OH$^+$ and H$_2$O$^+$ are missing. Ignoring 
Sgr B2, H$_3$O$^+$ is detected in only 6 of the components in which OH$^+$ and H$_2$O$^+$ 
are present. Apart from Sgr B2, OH$^+$ and H$_2$O$^+$ are found together in almost 
all velocity components. The ratio of column densities $N$(OH$^+$)/$N$(H$_2$O$^+$) 
typically lies in the range 1 - 10, with very few outliers.

Column densities for OH$^+$, H$_2$O$^+$, and atomic H are provided where possible by \citet{indr15}. 
These authors infer that the fraction of hydrogen as H$_2$ is generally small, $\sim$ a few percent. The fractional 
abundances of OH$^+$ typically lie in the range 10$^{-9}$ to $\sim$ 3 $\times$ 10$^{-8}$. 

Figures \ref{fig:3}a and \ref{fig:3}b show how the computed fractional abundances vary with number density 
and cosmic ray ionization rate. These fractional abundances are taken at the epoch at 
which peak OH$^+$ fractional abundance occurs. It is evident that low densities and high 
cosmic ray ionization rates give the largest fractional abundances, and 
that these are at the upper end of the observed range. As noted above, 
such low density clouds are unlikely to attain chemical steady-state 
within the lifetime of the clouds, and therefore peak fractional abundances are more 
appropriate to consider than steady-state values. Our results show that 
peak OH$^+$ fractional abundances in the observed range of about 1 $\times$ 10$^{-9}$ to 
3 $\times$ 10$^{-8}$ are obtained either with cloud 
number densities on the order of $\sim$ 10 cm$^{-3}$ and a canonical 
cosmic ray ionization rate, or in a cloud with a somewhat higher density, $\sim$ 100 cm$^{-3}$, but 
with a greatly enhanced cosmic ray rate, $\sim$ 50 $\times~ \zeta_{\text{ISM}}$. The computed H$_2$O$^+$ 
fractional abundance (taken at peak OH$^+$) is generally an order of magnitude less than that of OH$^+$ 
(consistent with observations), and H$_3$O$^+$ is even less abundant. Ratios OH$^+$/H$_2$O$^+$ closer to 
unity can be obtained when the chemistry is nearer steady-state, rather than at peak OH$^+$. However, 
these fractional abundances are low, in absolute terms, and a high cosmic ray flux would be 
required to raise them into the observed range. These results appear to be 
consistent with the observational data of \citet{indr15}. 
The solution for the O$^+$ hydrides based on very high cosmic ray ionization rates 
would also imply high fractional abundances of OH and H$_2$O (see Table \ref{tab:3}).

\subsection{Conclusions} 

While conventional gas-phase chemistry in diffuse clouds can account for the detected fractional 
abundances of hydrides of O$^+$, surface reactions contributing O atom 
hydrides to the gas make little difference to the O$^+$ ion fractional abundances and 
only marginal differences to the fractional abundances of neutrals OH and H$_2$O. The 
values of detected fractional abundances of the O$^+$ hydrides indicates that these 
species may be found either in conventional diffuse clouds ($n_{\text{H}}~\sim$ 
100 cm$^{-3}$) with very high cosmic ray ionization rates, or in very low density 
diffuse clouds with normal cosmic ray fluxes. 

On the other hand, the nitrogen hydrides detected in low density gas cannot be formed in the 
observed fractional abundances by gas-phase reactions, and surface reactions are 
required to form these hydrides, with an efficiency similar to that of H$_2$ 
formation. The location of these hydrides is indicated by these models to 
be fairly conventional diffuse clouds. The observational results \citep{per12} suggest 
that surface reactions inject not only NH$_3$ molecules but also NH$_2$ and NH radicals. 
Evidently, surface reactions may be important in diffuse cloud chemistry and their 
contribution in diffuse clouds gas-phase models should be assessed. However, hydride 
chemistry in dense dark clouds appears to be determined by gas-phase chemistry.

\section*{Acknowledgments}
We thank the referees for comments that helped to improve an earlier 
version of the manuscript. {\bf Z. Awad} thanks the Egyptian Science and Technology Development 
Funds, {\bf STDF}, for funding the research leading to these results through the (STF-Cycle 
5/2014-2015) under project ID: 12334. {\bf SV} acknowledges support from an STFC consolidated grant 
(grant number ST/M001334/1).


\begin{thebibliography}{}

\bibitem[{{Asplund} {et~al.}(2009){Asplund}, {Grevesse}, {Sauval}, \&
  {Scott}}]{asp09}
{Asplund}, M., {Grevesse}, N., {Sauval}, A.~J., \& {Scott}, P. 2009, \araa, 47,
  481

\bibitem[{{Bayet} {et~al.}(2011){Bayet}, {Williams}, {Hartquist}, \&
  {Viti}}]{bay11}
{Bayet}, E., {Williams}, D.~A., {Hartquist}, T.~W., \& {Viti}, S. 2011, \mnras,
  414, 1583

\bibitem[{{Crawford} \& {Williams}(1997)}]{craw97}
{Crawford}, I.~A., \& {Williams}, D.~A. 1997, \mnras, 291, L53

\bibitem[{{Dislaire} {et~al.}(2012){Dislaire}, {Hily-Blant}, {Faure}, {Maret},
  {Bacmann}, \& {Pineau Des For{\^e}ts}}]{disl12}
{Dislaire}, V., {Hily-Blant}, P., {Faure}, A., {et~al.} 2012, \aap, 537, A20

\bibitem[{{Draine}(2011)}]{dra11book}
{Draine}, B.~T. 2011, {Physics of the Interstellar and Intergalactic Medium}
  (Princeton University Press, Princeton and Oxford)

\bibitem[{{Fedoseev} {et~al.}(2015){Fedoseev}, {Ioppolo}, {Zhao}, {Lamberts},
  \& {Linnartz}}]{fed15}
{Fedoseev}, G., {Ioppolo}, S., {Zhao}, D., {Lamberts}, T., \& {Linnartz}, H.
  2015, \mnras, 446, 439

\bibitem[{{Garrod} {et~al.}(2008){Garrod}, {Weaver}, \& {Herbst}}]{garod08}
{Garrod}, R.~T., {Weaver}, S.~L.~W., \& {Herbst}, E. 2008, \apj, 682, 283

\bibitem[{{Gerin} {et~al.}(2010){Gerin}, {de Luca}, {Black}, {Goicoechea},
  {Herbst}, {Neufeld}, {Falgarone}, {Godard}, {Pearson}, {Lis}, {Phillips},
  {Bell}, {Sonnentrucker}, {Boulanger}, {Cernicharo}, {Coutens}, {Dartois},
  {Encrenaz}, {Giesen}, {Goldsmith}, {Gupta}, {Gry}, {Hennebelle},
  {Hily-Blant}, {Joblin}, {Kazmierczak}, {Kolos}, {Krelowski},
  {Martin-Pintado}, {Monje}, {Mookerjea}, {Perault}, {Persson}, {Plume},
  {Rimmer}, {Salez}, {Schmidt}, {Stutzki}, {Teyssier}, {Vastel}, {Yu},
  {Contursi}, {Menten}, {Geballe}, {Schlemmer}, {Shipman}, {Tielens},
  {Philipp-May}, {Cros}, {Zmuidzinas}, {Samoska}, {Klein}, \&
  {Lorenzani}}]{ger10}
{Gerin}, M., {de Luca}, M., {Black}, J., {et~al.} 2010, \aap, 518, L110

\bibitem[{{Gerlich}(1993)}]{gerl93}
{Gerlich}, D. 1993, J. Chem. Soc.{,} Faraday Trans., 89, 2199

\bibitem[{{Gupta} {et~al.}(2010){Gupta}, {Rimmer}, {Pearson}, {Yu}, {Herbst},
  {Harada}, {Bergin}, {Neufeld}, {Melnick}, {Bachiller}, {Baechtold}, {Bell},
  {Blake}, {Caux}, {Ceccarelli}, {Cernicharo}, {Chattopadhyay}, {Comito},
  {Cabrit}, {Crockett}, {Daniel}, {Falgarone}, {Diez-Gonzalez}, {Dubernet},
  {Erickson}, {Emprechtinger}, {Encrenaz}, {Gerin}, {Gill}, {Giesen},
  {Goicoechea}, {Goldsmith}, {Joblin}, {Johnstone}, {Langer}, {Larsson},
  {Latter}, {Lin}, {Lis}, {Liseau}, {Lord}, {Maiwald}, {Maret}, {Martin},
  {Martin-Pintado}, {Menten}, {Morris}, {M{\"u}ller}, {Murphy}, {Nordh},
  {Olberg}, {Ossenkopf}, {Pagani}, {P{\'e}rault}, {Phillips}, {Plume}, {Qin},
  {Salez}, {Samoska}, {Schilke}, {Schlecht}, {Schlemmer}, {Szczerba},
  {Stutzki}, {Trappe}, {van der Tak}, {Vastel}, {Wang}, {Yorke}, {Zmuidzinas},
  {Boogert}, {G{\"u}sten}, {Hartogh}, {Honingh}, {Karpov}, {Kooi}, {Krieg},
  {Schieder}, \& {Zaal}}]{gup10}
{Gupta}, H., {Rimmer}, P., {Pearson}, J.~C., {et~al.} 2010, \aap, 521, L47

\bibitem[{{Hidaka} {et~al.}(2011){Hidaka}, {Watanabe}, {Kouchi}, \&
  {Watanabe}}]{hid11}
{Hidaka}, H., {Watanabe}, M., {Kouchi}, A., \& {Watanabe}, N. 2011, Physical
  Chemistry Chemical Physics (Incorporating Faraday Transactions), 13, 15798

\bibitem[{{Hily-Blant} {et~al.}(2010){Hily-Blant}, {Maret}, {Bacmann},
  {Bottinelli}, {Parise}, {Caux}, {Faure}, {Bergin}, {Blake}, {Castets},
  {Ceccarelli}, {Cernicharo}, {Coutens}, {Crimier}, {Demyk}, {Dominik},
  {Gerin}, {Hennebelle}, {Henning}, {Kahane}, {Klotz}, {Melnick}, {Pagani},
  {Schilke}, {Vastel}, {Wakelam}, {Walters}, {Baudry}, {Bell}, {Benedettini},
  {Boogert}, {Cabrit}, {Caselli}, {Codella}, {Comito}, {Encrenaz}, {Falgarone},
  {Fuente}, {Goldsmith}, {Helmich}, {Herbst}, {Jacq}, {Kama}, {Langer},
  {Lefloch}, {Lis}, {Lord}, {Lorenzani}, {Neufeld}, {Nisini}, {Pacheco},
  {Phillips}, {Salez}, {Saraceno}, {Schuster}, {Tielens}, {van der Tak}, {van
  der Wiel}, {Viti}, {Wyrowski}, \& {Yorke}}]{hily10}
{Hily-Blant}, P., {Maret}, S., {Bacmann}, A., {et~al.} 2010, \aap, 521, L52

\bibitem[{{Hiraoka} {et~al.}(1995){Hiraoka}, {Yamashita}, {Yachi}, {Aruga},
  {Sato}, \& {Muto}}]{hir95}
{Hiraoka}, K., {Yamashita}, A., {Yachi}, Y., {et~al.} 1995, \apj, 443, 363

\bibitem[{{Hollenbach} {et~al.}(2012){Hollenbach}, {Kaufman}, {Neufeld},
  {Wolfire}, \& {Goicoechea}}]{holn12}
{Hollenbach}, D., {Kaufman}, M.~J., {Neufeld}, D., {Wolfire}, M., \&
  {Goicoechea}, J.~R. 2012, \apj, 754, 105

\bibitem[{{Indriolo} {et~al.}(2015){Indriolo}, {Neufeld}, {Gerin}, {Schilke},
  {Benz}, {Winkel}, {Menten}, {Chambers}, {Black}, {Bruderer}, {Falgarone},
  {Godard}, {Goicoechea}, {Gupta}, {Lis}, {Ossenkopf}, {Persson},
  {Sonnentrucker}, {van der Tak}, {van Dishoeck}, {Wolfire}, \&
  {Wyrowski}}]{indr15}
{Indriolo}, N., {Neufeld}, D.~A., {Gerin}, M., {et~al.} 2015, \apj, 800, 40

\bibitem[{{Jura}(1974)}]{jura74}
{Jura}, M. 1974, \apj, 191, 375

\bibitem[{{Knauth} {et~al.}(2004){Knauth}, {Andersson}, {McCandliss}, \&
  {Warren Moos}}]{kna04}
{Knauth}, D.~C., {Andersson}, B.-G., {McCandliss}, S.~R., \& {Warren Moos}, H.
  2004, \nat, 429, 636

\bibitem[{{Le Bourlot}(1991)}]{lebo91}
{Le Bourlot}, J. 1991, \aap, 242, 235

\bibitem[{{Le Gal} {et~al.}(2014b){Le Gal}, {Hily-Blant}, \&
  {Faure}}]{legal14sf}
{Le Gal}, R., {Hily-Blant}, P., \& {Faure}, A. 2014b, in SF2A-2014: Proceedings
  of the Annual meeting of the French Society of Astronomy and Astrophysics,
  ed. J.~{Ballet}, F.~{Martins}, F.~{Bournaud}, R.~{Monier}, \& C.~{Reyl{\'e}},
  397--401

\bibitem[{{Le Gal} {et~al.}(2014a){Le Gal}, {Hily-Blant}, {Faure}, {Pineau des
  For{\^e}ts}, {Rist}, \& {Maret}}]{legal14}
{Le Gal}, R., {Hily-Blant}, P., {Faure}, A., {et~al.} 2014a, \aap, 562, A83

\bibitem[{{Le Petit} {et~al.}(2004){Le Petit}, {Roueff}, \& {Herbst}}]{lepet04}
{Le Petit}, F., {Roueff}, E., \& {Herbst}, E. 2004, \aap, 417, 993

\bibitem[{{Le Teuff} {et~al.}(2000){Le Teuff}, {Millar}, \&
  {Markwick}}]{lete2000}
{Le Teuff}, Y.~H., {Millar}, T.~J., \& {Markwick}, A.~J. 2000, \aaps, 146, 157

\bibitem[{{Linnartz} {et~al.}(2015){Linnartz}, {Ioppolo}, \&
  {Fedoseev}}]{lin15}
{Linnartz}, H., {Ioppolo}, S., \& {Fedoseev}, G. 2015, ArXiv e-prints,
  arXiv:1507.02729

\bibitem[{{Marquette} {et~al.}(1988){Marquette}, {Rebrion}, \& {Rowe}}]{marq88}
{Marquette}, J.~B., {Rebrion}, C., \& {Rowe}, B.~R. 1988, \jcp, 89, 2041

\bibitem[{{McCrea} \& {McNally}(1960)}]{mccr60}
{McCrea}, W.~H., \& {McNally}, D. 1960, \mnras, 121, 238

\bibitem[{{McElroy} {et~al.}(2013){McElroy}, {Walsh}, {Markwick}, {Cordiner},
  {Smith}, \& {Millar}}]{mce013}
{McElroy}, D., {Walsh}, C., {Markwick}, A.~J., {et~al.} 2013, \aap, 550, A36

\bibitem[{{Meyer} \& {Roth}(1991)}]{mey91}
{Meyer}, D.~M., \& {Roth}, K.~C. 1991, \apjl, 376, L49

\bibitem[{{Millar} {et~al.}(1997){Millar}, {Farquhar}, \& {Willacy}}]{milr97}
{Millar}, T.~J., {Farquhar}, P.~R.~A., \& {Willacy}, K. 1997, \aaps, 121, 139

\bibitem[{{Persson} {et~al.}(2010){Persson}, {Black}, {Cernicharo},
  {Goicoechea}, {Hassel}, {Herbst}, {Gerin}, {de Luca}, {Bell}, {Coutens},
  {Falgarone}, {Goldsmith}, {Gupta}, {Ka{\'z}mierczak}, {Lis}, {Mookerjea},
  {Neufeld}, {Pearson}, {Phillips}, {Sonnentrucker}, {Stutzki}, {Vastel}, {Yu},
  {Boulanger}, {Dartois}, {Encrenaz}, {Geballe}, {Giesen}, {Godard}, {Gry},
  {Hennebelle}, {Hily-Blant}, {Joblin}, {Ko{\l}os}, {Kre{\l}owski},
  {Mart{\'{\i}}n-Pintado}, {Menten}, {Monje}, {Perault}, {Plume}, {Salez},
  {Schlemmer}, {Schmidt}, {Teyssier}, {P{\'e}ron}, {Cais}, {Gaufre}, {Cros},
  {Ravera}, {Morris}, {Lord}, \& {Planesas}}]{per10}
{Persson}, C.~M., {Black}, J.~H., {Cernicharo}, J., {et~al.} 2010, \aap, 521,
  L45

\bibitem[{{Persson} {et~al.}(2012){Persson}, {De Luca}, {Mookerjea},
  {Olofsson}, {Black}, {Gerin}, {Herbst}, {Bell}, {Coutens}, {Godard},
  {Goicoechea}, {Hassel}, {Hily-Blant}, {Menten}, {M{\"u}ller}, {Pearson}, \&
  {Yu}}]{per12}
{Persson}, C.~M., {De Luca}, M., {Mookerjea}, B., {et~al.} 2012, \aap, 543,
  A145

\bibitem[{{Pickles} \& {Williams}(1977)}]{pic77}
{Pickles}, J.~B., \& {Williams}, D.~A. 1977, \apss, 52, 443

\bibitem[{{Roberts} {et~al.}(2007){Roberts}, {Rawlings}, {Viti}, \&
  {Williams}}]{rob07}
{Roberts}, J.~F., {Rawlings}, J.~M.~C., {Viti}, S., \& {Williams}, D.~A. 2007,
  \mnras, 382, 733

\bibitem[{{Tielens}(2005)}]{tiel05}
{Tielens}, A.~G.~G.~M. 2005, {The Physics and Chemistry of the Interstellar
  Medium} (Cambridge University Press,~Cambridge, UK)

\bibitem[{{Viti} {et~al.}(2004){Viti}, {Collings}, {Dever}, {McCoustra}, \&
  {Williams}}]{viti04}
{Viti}, S., {Collings}, M.~P., {Dever}, J.~W., {McCoustra}, M.~R.~S., \&
  {Williams}, D.~A. 2004, \mnras, 354, 1141

\bibitem[{{Viti} \& {Williams}(1999)}]{viti99}
{Viti}, S., \& {Williams}, D.~A. 1999, \mnras, 305, 755

\bibitem[{{Wagenblast} \& {Williams}(1996)}]{wag96}
{Wagenblast}, R., \& {Williams}, D.~A. 1996, \apss, 236, 257

\bibitem[{{Wagenblast} {et~al.}(1993){Wagenblast}, {Williams}, {Millar}, \&
  {Nejad}}]{wag93MN}
{Wagenblast}, R., {Williams}, D.~A., {Millar}, T.~J., \& {Nejad}, L.~A.~M.
  1993, \mnras, 260, 420

\bibitem[{{Wakelam} {et~al.}(2005){Wakelam}, {Selsis}, {Herbst}, \&
  {Caselli}}]{wak05}
{Wakelam}, V., {Selsis}, F., {Herbst}, E., \& {Caselli}, P. 2005, \aap, 444,
  883

\bibitem[{{Woodall} {et~al.}(2007){Woodall}, {Ag{\'u}ndez}, {Markwick-Kemper},
  \& {Millar}}]{woo07}
{Woodall}, J., {Ag{\'u}ndez}, M., {Markwick-Kemper}, A.~J., \& {Millar}, T.~J.
  2007, \aap, 466, 1197

\bibitem[{{Zymak} {et~al.}(2013){Zymak}, {Hejduk}, {Mulin}, {Pla{\v s}il},
  {Glos{\'{\i}}k}, \& {Gerlich}}]{zym13}
{Zymak}, I., {Hejduk}, M., {Mulin}, D., {et~al.} 2013, \apj, 768, 86

\end{thebibliography}

\begin{table*}
   \centering
\caption {The main gas-phase entry chemical reactions in our network to form N-, O- and O$^+$-hydrides with their 
rate constants taken from UMIST 2012 ratefile \citep{mce013}. The rate constants (k, in cm$^{3}$ s$^{-1}$) are calculated 
from $k = \alpha~(T/300)^{\beta}~exp(-\gamma/T)$.}
  \label{tab:1}
\scalebox{0.9}
{
\begin{tabular}{lclclclcllrr} \\ \hline \hline
\multicolumn{9}{c}{\bf Chemical Reactions}& {\bf $\alpha$}& {\bf $\beta$} & {\bf $\gamma$}\\ \hline \\
H$_2$  &  +  &  CR       & $\rightarrow$  &  H$_2^+$     &  +  &  e$^-$&    &     & 1.20E-17 & 0.00  &  0.0\\[0.5ex]
H$_2^+$ &  +  &  H$_2$    & $\rightarrow$  &  H$_3^+$     &  +  &   H  &    &     & 2.08E-09 & 0.00  &  0.0\\[0.5ex]\\

H      & + & CR   & $\rightarrow$  &  H$^+$  &  +  &e$^-$  &  &  & 5.98E-18 & 0.00  &  0.0\\[0.5ex]
H$^+$  & + & O    &  $\rightarrow$ &  O$^+$  &  +  & H    &  &  & 6.86E-10 & 0.26  &  224.3\\[0.5ex]
H      & + & O$^+$&  $\rightarrow$ &  O      &  +  & H$^+$&  &  & 5.66E-10 & 0.36  &  -8.6\\[0.5ex]
H$_2$  & + & O$^+$&  $\rightarrow$ &  OH$^+$ &  +  & H    &  &  & 1.70E-09 & 0.00  &  0.0\\[0.5ex]
H$_3^+$& + & O    &  $\rightarrow$ &  OH$^+$ &  +  & H$_2$&  &  & 7.98E-10 & -0.16 &  1.4\\[0.5ex]
OH$^+$ & + &e$^-$  & $\rightarrow$  &  O      &  +  & H   &  &  & 3.75E-08 & -0.50 &  0.0\\[0.5ex]
H$_2$  &  +  &  OH$^+$   &  $\rightarrow$ &  H$_2$O$^+$ &  +  &   H &    &      &  1.01E-09  &  0.00  &  0.0\\[0.5ex]
H$_3^+$&  +  &  O        &  $\rightarrow$ &  H$_2$O$^+$ &  +  &   H &    &      &  3.42E-10  &  -0.16  &  1.4\\[0.5ex]
H$_2$O$^+$ &+&e$^-$ &  $\rightarrow$  & OH  &  +  & H     & & & 8.60E-08 & -0.50 & 0.0 \\[0.5ex]
H$_2$  &  +  &H$_2$O$^+$ &  $\rightarrow$ &  H$_3$O$^+$ &  +  &   H &    &      &  6.40E-10  &  0.00  &  0.0\\[0.5ex]
H$_3$O$^+$ &+&e$^-$ &  $\rightarrow$  & H$_2$O&+  & H     & & & 7.09E-08 & -0.50 & 0.0 \\[0.5ex]
H$_3$O$^+$ &+&e$^-$ &  $\rightarrow$  & OH  &  +  & H$_2$ & & & 5.37E-08 & -0.50 & 0.0 \\[0.5ex]
H$_3$O$^+$ &+&e$^-$ &  $\rightarrow$  & OH  &  +  & H     &+&H& 3.05E-07 & -0.50 & 0.0 \\[0.5ex]\\

N        &  +  & CR   &  $\rightarrow$  &  N$^+$    &  +  &e$^-$  &     &         & 2.70E-17&  0.00&      0.0\\[0.5ex]
N$^+$    &  +  & H$_2$&  $\rightarrow$  &  NH$^+$   &  +  & H   &     &         & 1.00E-09&  0.00&     85.0\\[0.5ex]
NH$^+$   &  +  & H$_2$&  $\rightarrow$  &  NH$_2^+$ &  +  & H   &     &         & 1.28E-09&  0.00&      0.0\\[0.5ex]
NH$_2^+$ &  +  &e$^-$  &  $\rightarrow$  &  NH       &  +  & H   &     &         & 9.21E-08& -0.79&     17.1\\[0.5ex]
NH$_2^+$ &  +  & H$_2$&  $\rightarrow$  &  NH$_3^+$ &  +  & H   &     &         & 2.70E-10&  0.00&      0.0\\[0.5ex]
NH$_3^+$ &  +  &e$^-$  &  $\rightarrow$  &  NH       &  +  & H   &  +  &   H    & 1.55E-07& -0.50&      0.0\\[0.5ex]
NH$_3^+$ &  +  &e$^-$  &  $\rightarrow$  &  NH$_2$   &  +  & H   &     &         & 1.55E-07& -0.50&      0.0\\[0.5ex]
NH$_3^+$ &  +  & H$_2$&  $\rightarrow$  &  NH$_4^+$ &  +  & H   &     &         & 3.09E-13&  1.08&    -50.9\\[0.5ex]
NH$_4^+$ &  +  &e$^-$  &  $\rightarrow$  &  NH$_2$   &  +  & H$_2$&    &         & 4.72E-08& -0.60&      0.0\\[0.5ex]
NH$_4^+$ &  +  &e$^-$  &  $\rightarrow$  &  NH$_2$   &  +  & H   &  +  &   H    & 3.77E-08& -0.60&      0.0\\[0.5ex]
NH$_4^+$ &  +  &e$^-$  &  $\rightarrow$  &  NH$_3$   &  +  & H   &     &         & 8.49E-07& -0.60&      0.0\\[0.5ex]
NH$_3$   &  +  & h+   &  $\rightarrow$  &  NH$_3^+$ &  +  & H   &     &         & 3.70E-09& -0.50&      0.0\\[0.5ex]
NH$_3$   &  +  & H$_3^+$&  $\rightarrow$&  NH$_4^+$ &  +  & H$_2$&     &         & 4.39E-09& -0.50&      0.0\\[0.5ex]
N$_2$    &  +  & H$_3^+$&  $\rightarrow$& N$_2$H$^+$&  +  & H$_2$&     &         & 1.80E-09&  0.00&      0.0\\[0.5ex]
N$_2$H$^+$&  +  &e$^-$   & $\rightarrow$ &  N        &  +  & NH   &     &         & 2.09E-08& -0.74&      0.0\\[0.5ex]
N$_2$     &  +  & He$^+$   & $\rightarrow$ &  N$^+$ &  +  & N    &  +  &   He    & 9.60E-10&  0.00&      0.0\\ \hline \hline
\end{tabular} 
}
\end{table*}
\begin{table*}
   \centering
\caption {The fractional abundances and ratios at different number density and radiation field, for N-hydrides.}
  \label{tab:2}
\scalebox{0.9}
{
\begin{tabular}{ccccccc} \\ \hline \hline
{\bf Number density} & {\bf RF}& \multicolumn{3}{c}{\bf $^\dag$Fractional Abundances}& \multicolumn{2}{c}{\bf Ratio}\\[0.5ex]
{\bf (cm$^{-3}$)}& {\bf(G$_0$)}&{\bf x(NH)} & {\bf x(NH$_2$)}&{\bf x(NH$_3$)} & {\bf x(NH$_3$)/x(NH)}& {\bf x(NH$_3$)/x(NH$_2$)} \\[0.5ex] \hline \\
  & 0.25 & 7.90$\times$10$^{-10}$ & 2.65$\times$10$^{-10}$ & 2.36$\times$10$^{-10}$ & 0.29 & 0.89\\
 3 & 0.5 & 4.443$\times$10$^{-10}$ & 1.416$\times$10$^{-10}$ & 1.223$\times$10$^{-10}$ & 0.28 & 0.86\\
  & 1 & 2.360$\times$10$^{-10}$ & 7.306$\times$10$^{-11}$ & 6.226$\times$10$^{-11}$ & 0.26 & 0.85 \\ [0.5ex] \hline \\  
  & 0.25  & 1.526$\times$10$^{-9}$  & 6.844$\times$10$^{-10}$ & 7.616$\times$10$^{-10}$ & 0.5 & 1.1 \\
 10 & 0.5 & 1.004$\times$10$^{-9}$  & 3.987$\times$10$^{-10}$ & 4.008$\times$10$^{-10}$ & 0.4 & 1.005\\
  & 1     & 5.912$\times$10$^{-10}$ & 2.172$\times$10$^{-10}$ & 2.059$\times$10$^{-10}$ & 0.35 & 0.95 \\[0.5ex] \hline \\
  & 0.25  & 2.049$\times$10$^{-9}$ & 1.276$\times$10$^{-9}$  & 1.983$\times$10$^{-9}$ & 0.9 & 1.5\\
 30 & 0.5 & 1.690$\times$10$^{-9}$ & 8.771$\times$10$^{-10}$ & 1.111$\times$10$^{-9}$ & 0.66 & 1.26\\
    & 1   & 1.213$\times$10$^{-9}$ & 5.38$\times$10$^{-10}$  & 5.925$\times$10$^{-10}$ & 0.49 & 1.1 \\[0.5ex] \hline \\
  & 0.25   & 2.259$\times$10$^{-9}$ & 1.855$\times$10$^{-9}$ & 4.665$\times$10$^{-9}$ & 2.1  & 2.5 \\
 100 & 0.5 & 2.099$\times$10$^{-9}$ & 1.540$\times$10$^{-9}$ & 2.960$\times$10$^{-9}$ & 1.4  & 1.9 \\
     & 1   & 1.895$\times$10$^{-9}$ & 1.161$\times$10$^{-9}$ & 1.737$\times$10$^{-9}$ & 0.92 & 1.5 \\[0.5ex] \hline \hline
\end{tabular} 
}
\flushleft {$\dag$ The fractional abundances, x(Y), are defined by $n$(Y)/$n_{\text{H}}$.}
\end{table*}

\begin{table*}
\centering
\caption{Fractional abundances and ratios for the RM, for the case in which surface reactions 
provide NH, NH$_2$, and NH$_3$ either in equal proportions (Model A), or in varying proportions, 0.5, 0.3, 0.2, 
respectively, (Model B).} 
\label{tab:4}
\scalebox{0.9}
{
\begin{tabular}{cccccccc} \\\hline \hline
{\bf Density} & {\bf RF} &{\bf Model} & \multicolumn{3}{c}{\bf $\dag$Fractional Abundances}& \multicolumn{2}{c}{\bf Ratio}\\[0.5ex]
{\bf (cm$^{-3}$)}& {\bf(G$_0$)}&&{\bf x(NH)} & {\bf x(NH$_2$)}&{\bf x(NH$_3$)} &{\bf x(NH$_3$)/x(NH)}&{\bf x(NH$_3$)/x(NH$_2$)}\\ \hline \\
 100 & 0.25& A & 2.654$\times$10$^{-9}$ & 1.631$\times$10$^{-9}$ & 1.439$\times$10$^{-9}$ & 0.54 & 0.88 \\
     &     & B & 3.120$\times$10$^{-9}$ & 1.307$\times$10$^{-9}$ & 9.332$\times$10$^{-10}$& 0.29 & 0.71 \\ [0.5ex]\hline \\
  200 & 0.25& A & 2.967$\times$10$^{-9}$ & 1.967$\times$10$^{-9}$ & 2.283$\times$10$^{-9}$ & 0.75 & 1.14 \\
      &     & B & 3.490$\times$10$^{-9}$ & 1.582$\times$10$^{-9}$ & 1.384$\times$10$^{-9}$ & 0.39 & 0.87 \\\hline \hline
\end{tabular} 
}
\flushleft {$\dag$ The fractional abundances, x(Y), are defined by $n$(Y)/$n_{\text{H}}$.}
\end{table*}
\begin{table*}
   \centering
\caption {The fractional abundances and ratios at different number density and CR ionization rate for O-, and O$^+$- hydrides. 
Note that the results are shown at the peak position of OH$^+$,`{\bf p}', and at steady-state, `{\bf ss}'.}
  \label{tab:3}
\scalebox{0.9}
{
\begin{tabular}{ccccccc} \\ \hline \hline
{\bf Number density} & {\bf $\zeta / \zeta_{\text{ISM}}$} & \multicolumn{4}{c}{\bf Fractional Abundances} &{\bf Ratio}\\ [0.5ex]
{\bf (cm$^{-3}$)}     &    & {\bf x(OH)} & {\bf x(H$_2$O)}&{\bf x(OH$^+$)}  &{\bf x(H$_2$O$^+$)} & {\bf x(OH$^+$)/ x(H$_2$O$^+$)}\\ \hline \\
  100 & 50 & 1.786 $\times$10$^{-7}$  & 4.677 $\times$10$^{-9}$ & 2.397$\times$10$^{-8}$ & 2.589 $\times$10$^{-9}$  & 9.258 {\bf $^{\text p}$}   \\
      &  & 5.736 $\times$10$^{-7}$  & 3.743 $\times$10$^{-8}$ &1.711$\times$10$^{-9}$ & 1.435$\times$10$^{-9}$  & 1.192 {\bf $^{\text{ss}}$}  \\ [0.5ex] \hline \\ 
 10 & 1.0 & 2.767 $\times$10$^{-8}$  & 2.085 $\times$10$^{-9}$ & 8.074$\times$10$^{-9}$  & 8.732 $\times$10$^{-10}$  & 9.246 {\bf $^{\text p}$} \\
    &   & 3.089 $\times$10$^{-8}$  & 3.770 $\times$10$^{-9}$ & 2.732 $\times$10$^{-10}$ & 2.374 $\times$10$^{-10}$ & 1.151 {\bf $^{\text{ss}}$} \\ \hline \hline \\
\end{tabular} 
}
\flushleft {$\dag$ The fractional abundances, x(Y), are defined by $n$(Y)/$n_{\text{H}}$.}
\end{table*}

\end{document}